\documentclass[a4paper,twocolumn,fleqn]{article} 
\usepackage{txfonts}                   
\usepackage[dvipdfmx]{graphicx}       

\usepackage{tabularx}
\usepackage{pfr_APiP2025}             
\usepackage{siunitx}

\begin{document}

\title{Photoexcitation spectroscopy of highly charged ions for application to astronomy using a compact electron beam ion trap (EBIT) \\at the synchrotron radiation facility SPring-8}

\author{Leo~HIRATA\sup{1,2,3}, Yuki~AMANO\sup{2,3}, Moto~TOGAWA\sup{4,5}, Hiroyuki~A.~SAKAUE\sup{6}, Nobuyuki~NAKAMURA\sup{7,3}, Makoto~SAWADA\sup{8,3}, Hiromasa~SUZUKI\sup{9,3}, Masaki~OURA\sup{10} and Hiroya~YAMAGUCHI\sup{2,1,3}}

\affiliation{\normalsize 
  \sup{1}Department of Physics, The University of Tokyo, 7-3-1 Hongo, Bunkyo-ku, Tokyo 113-0033, Japan \\
  \sup{2}Institute of Space and Astronautical Science (ISAS), Japan Aerospace Exploration Agency (JAXA), 3-1-1 Yoshinodai, Chuo-ku, Sagamihara 252-5210, Japan \\
  \sup{3}RIKEN Pioneering Research Institute, RIKEN, 2-1 Hirosawa, Wako 351-0198, Japan \\
  \sup{4}Max-Planck-Institut für Kernphysik, Saupfercheckweg 1, 69117 Heidelberg, Germany \\
  \sup{5}European XFEL, Holzkoppel 4, 22869 Schenefeld, Germany \\
  \sup{6}National Institute for Fusion Science, National Institutes of Natural Sciences, 322-6 Oroshi-cho, Toki 509-5292, Japan \\
  \sup{7}Institute for Laser Science, The University of Electro-Communications, 1-5-1 Chofugaoka, Chofu 182-8585, Japan \\
  \sup{8}Department of Physics, Rikkyo University, 3-34-1 Nishi Ikebukuro, Toshima-ku, Tokyo 171-8501, Japan \\
  \sup{9}Faculty of Engineering, University of Miyazaki, 1-1 Gakuen Kibanadai-nishi, Miyazaki 889-2192 Japan \\
  \sup{10}RIKEN SPring-8 Center, RIKEN, 1-1-1 Kouto, Sayo, Hyogo 679-5148, Japan}

\date{\small January 26, 2026}

\email{lhirata-6087@g.ecc.u-tokyo.ac.jp}

\begin{abstract}
In the past few decades, X-ray astronomy satellites equipped with grating spectrometers and microcalorimeters have enabled high-resolution spectroscopic observations of astrophysical objects. The need for accurate atomic data has arose as we attempt detailed analysis of the high-resolution spectra they provide. This is because current spectral models, which heavily rely on theoretical calculations, entail non-negligible uncertainties.
We employ a plasma spectroscopy device called electron beam ion trap (EBIT) to experimentally obtain precise atomic data. An EBIT with a design that allows combined operation with synchrotron radiation facilities was developed based on the Heidelberg Compact EBIT and installed at ISAS/JAXA for this purpose. 
We conducted a spectroscopic experiment using the JAXA-EBIT at the synchrotron radiation facility SPring-8, and successfully obtained high-resolution spectra of the L$\alpha$ resonance transition of Ne-like Fe$^{16+}$ ions, 3C, as well as the K$\alpha$ resonance transition of He-like O$^{6+}$ ions.
We also measured another Ne-like Fe$^{16+}$ L$\alpha$ resonance transition, 3G, and constrained an upper limit of the oscillator strength ratio of 3G to 3C, using our experimental results.
The experimental values obtained in this study will be applied to observational studies of astrophysical objects as a part of the plasma spectral modeling. 

\end{abstract}


\AbstNum{B-13}

\maketitle

\begin{normalsize}

\newpage
\section{Introduction}

Various astrophysical objects, including supernova remnants and galaxy clusters, consist of X-ray-emitting hot ($\sim 10^{6-8}$ K) plasma (e.g., \cite{Hitomi16}). 
XRISM \cite{Tashiro18} satellite’s spectrometer, Resolve \cite{Ishisaki22}, provides exceptional spectral resolution $\Delta E/E \sim 8 \times 10^{-4}$ at 6 keV 
\cite{XRISM24}, resolving the fine iron K-shell emission line structures (e.g., \cite{XRISM25}). 
Such high resolving power enables us to perform precise diagnostics of astrophysical plasmas, including detailed measurements of velocity structures \cite{XRISM25b, Suzuki25} and elemental abundances \cite{Plucinsky25} of the astrophysical objects. 
The observed spectra are often analysed by comparing them with theoretically synthesized spectra using plasma modeling frameworks such as AtomDB/APEC \cite{Smith01, Foster12}, SPEX \cite{dePlaa20}, and CHIANTI \cite{Dere97, Zanna15}.
The accuracy of the atomic data used in those frameworks is crucial for reliable plasma diagnostics using high-resolution spectra. 
However, currently used atomic data, which heavily rely on theoretical calculations, entail non-negligible uncertainties up to 30-40\% depending on the transitions \cite{Hitomi18}.
Laboratory measurements are necessary to obtain precise values.
To address this issue, we considered the use of an electron beam ion trap (EBIT) \cite{Marrs88}, an experimental apparatus designed to produce and trap highly charged ions (HCIs) using a monochromatic electron beam and to measure their X-ray emission. 

Since the development of the EBIT at Lawrence Livermore National Laboratory (LLNL) in the 1980s, various experiments to measure atomic data have been carried out through spectroscopic observations of EBIT plasmas using different types of EBITs operated at, for example, LLNL, National Institute of Standards and Technology (NIST), and Max-Planck-Institut für Kernphysik (MPIK). These experiments have targeted a variety of transitions in elements of astrophysical importance, including oxygen \cite{Leutenegger20}, silicon \cite{Baumann14, Hell16}, sulfur \cite{Hell16, Rahin25}, argon \cite{Beilmann13, Gall19}, and iron \cite{Beiersdorfer02, Shah25}. Thus far, such experiments have been conducted primarily with EBITs alone, where the emission from EBIT plasmas induced by electron-impact excitation and recombination was observed with crystal or grating spectrometers and microcalorimeters, allowing precise determinations of, for instance, line intensity ratios, centroid energies, and excitation cross sections, through passive spectroscopy (e.g., \cite{Beiersdorfer02, Rahin25, Shah21}).
In recent years, the development of EBITs designed to allow synchrotron radiation to be injected into the trapped plasma, such as the FLASH-EBIT \cite{Epp10} and the Heidelberg Compact EBIT (HC-EBIT) \cite{Micke18}, has enabled the reproduction of transitions driven by photon–ion interactions, such as resonant photoexcitation processes (e.g., \cite{Kuhn20, Steinbrugge22}) (Fig. \ref{fig:EBIT_beamline_schematic}). These processes are of crucial importance in astrophysical plasmas but could not be realized with EBITs alone. Moreover, by exploiting the high monochromaticity of synchrotron radiation, active spectroscopy—performed by scanning the incident photon energy—has become possible, providing excellent energy resolution surpassing that of existing X-ray spectrometers. Kühn \textit{et al.} \cite{Kuhn22} has successfully obtained the oscillator strength ratio of two L-shell transition lines (referred to as 3C and 3D) of highly charged iron ions, with a systematic uncertainty of $\approx$ 2\%, using this technique. 

\begin{figure*}[ht]
  \centering
  \includegraphics[width=2.0\columnwidth]{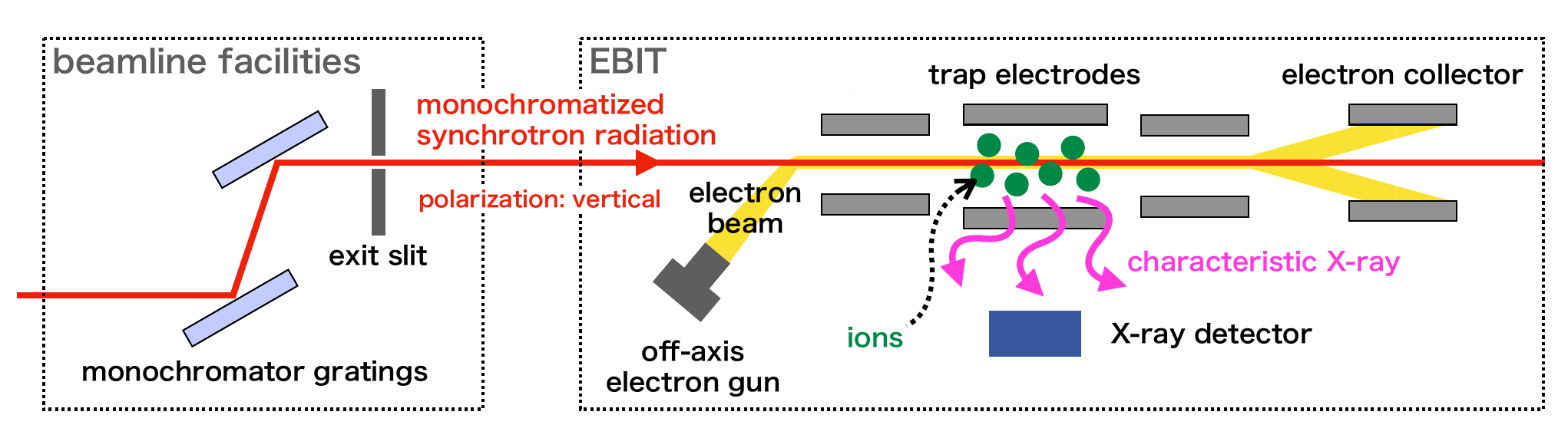}
  \caption{Schematic diagram showing the mechanism of HCI production and trapping inside the JAXA-EBIT, as well as the detection of the photons emitted from the resonant photoexcitation processes of HCIs irradiated with the monochromatic X-ray photons of the synchrotron radiation, which is introduced coaxially through the EBIT drift tube.}
  \label{fig:EBIT_beamline_schematic}
\end{figure*}

With the aim of making such an EBIT-based experimental technique capable of high-accuracy atomic data measurement for a wide variety of elements more accessible to the Japanese astronomical community, we developed, in collaboration with MPIK, a compact EBIT based on the HC-EBIT design, which can be operated in conjunction with synchrotron radiation facilities. This device, hereafter the JAXA-EBIT, was completed and installed at ISAS/JAXA in 2022.
The details of the specifications and performance of our device are reported in \cite{Amano25}.
As a first step of comprehensive atomic data measurement using the JAXA-EBIT, we focus on the L-shell transition lines of highly charged iron ions (also known as the Fe-L complex). 
The lines in the Fe-L complex are frequently highlighted in observational studies of galaxy clusters (e.g., \cite{Werner09, dePlaa12}) and supernova remnants (e.g., \cite{Hughes03, Amano20}). 
In particular, we examine the resonance lines from the $\lbrack 2p^5_{1/2} 3d_{3/2} \rbrack_{J=1} \rightarrow \lbrack 2p^6 \rbrack_{J=0}$ (3C) and $\lbrack 2p^5_{3/2} 3s_{1/2} \rbrack_{J=1} \rightarrow \lbrack 2p^6 \rbrack_{J=0}$ (3G) transitions in Ne-like Fe$^{16+}$ ions. These lines are one of the most prominent spectral features among the Fe-L complex that are useful in measuring various types of astrophysical properties thanks to their excellent photon statistics. 
However, the complexity of the multi-electron system obstructs precise theoretical calculations, and the intensity ratio of these two lines has been known to show discrepancies between observational, experimental, and theoretical values \cite{LGu19, Shah19}. 
Such discrepancies are considered to result from the complicated effect of electron collision, the quality of fundamental wave functions, and the accuracy of atomic data, such as oscillator strengths. 
A high-resolution photoexcitation spectroscopy of HCIs, using the EBIT and synchrotron radiation together, can provide precise measurements of the centroid energies and oscillator strengths of these transitions. 

\section{Experiment}

We aim to precisely measure the centroid energies as well as intensities of Ne-like Fe$^{16+}$ 3C and 3G lines, and experimentally constrain the oscillator strength ratio of 3G to 3C (i.e., $f_\textrm{3G}/f_\textrm{3C}$) within the precision of 10\%; a typical precision requirement for observational studies of astrophysical objects. 
Note that theoretically calculated oscillator strengths of Ne-like Fe$^{16+}$ 3C and 3G lines, provided in AtomDB based on \cite{Loch06}, are 2.49 and 0.126, respectively.
Therefore, we brought the JAXA-EBIT to the soft X-ray beamline BL17SU \cite{Ohashi07} at the synchrotron radiation facility SPring-8 \cite{Hara00} and conducted a six-day beam-time experiment to perform active spectroscopy on the resonant photoexcitation processes of EBIT plasma. 

\begin{figure*}[ht]
  \centering
  \includegraphics[width=2.0\columnwidth]{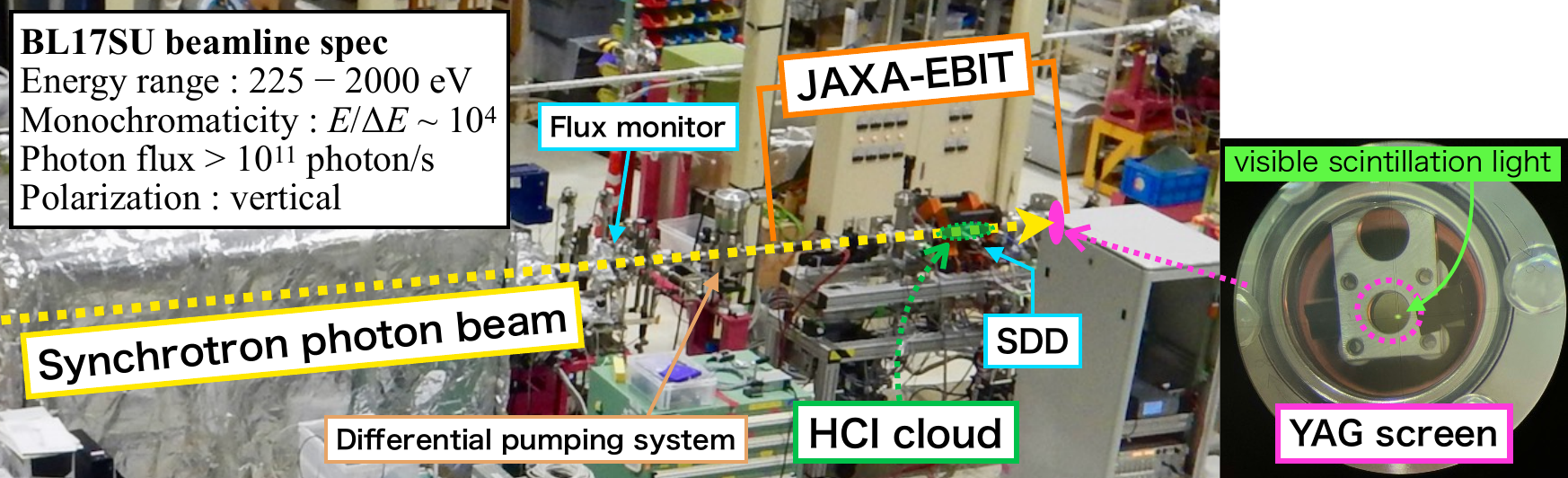}
  \caption{Photograph of the JAXA-EBIT connected to BL17SU, SPring-8, with key components in our experimental setup.}
  \label{fig:SP8_beamline_setup_photo}
\end{figure*}

\subsection{Experimental setup and technique}
A photograph of the experimental setup is shown in Figure \ref{fig:SP8_beamline_setup_photo}, and a schematic diagram of this experiment, along with main components of the setup, is presented in Figure \ref{fig:EBIT_beamline_schematic}.
At the soft X-ray beamline BL17SU, the synchrotron radiation beam emitted from the helical-8 undulator (ID17) \cite{Tanaka23} is monochromatized by a grating monochromator \cite{Ohashi07, Senba07}.
BL17SU provides a photon beam with high monochromaticity ($\Delta E/E \sim 1 \times 10^{-4}$) and flux ($\sim 10^{11}$ photon/s) according to the nominal performances.
The incident photon energy was calibrated prior to the beam-time by conducting X-ray photoemission spectroscopy (XPS) measurement of 4\textit{f} electrons from the thin Au foil using an XPS apparatus, which was located downstream of our EBIT. 
A Kirkpatrick Baez (KB) mirror \cite{KBmirror} system refocuses the photon beam at the position of the JAXA-EBIT trap region a few meters downstream. 

We connected the JAXA-EBIT to BL17SU branch-a through a differential pumping system. 
Monochromatized synchrotron radiation was irradiated onto the HCIs produced in the EBIT, and X-ray photons were resonantly absorbed, inducing resonant photoexcitation, and re-emitted by the HCIs. 
These photons were detected by a silicon drift detector (SDD), equipped with a 500 nm thin aluminum filter, mounted on the side of the central trap region. KETEK VITUS H150 SDD was employed in our experimental setup. VITUS H150's circular detection surface is 13.6 mm in diameter, with a collimated area of 143 mm$^2$ and an absorption depth of 450 \unit{\micro\meter}. Its operating temperature is typically around 228 K. 
To efficiently detect the X-ray photons emitted by resonant photoexcitation processes, the SDD was mounted perpendicular to the beam axis and the polarization direction of the incident photon beam; in this experiment we only used the vertically polarized photon beam. 
We change the incident photon energy step by step to search for resonance events. (Fig. \ref{fig:sync_ener_scan_diagram}; hereafter we call this method a scan.)
By measuring the dependence of detected photon counts of each line on the incident photon energy, we can obtain high-resolution spectra, utilizing the excellent monochromaticity of the synchrotron photon beam.
We also installed an Au foil photoelectron meter that can be inserted onto the beam axis when needed upstream of the JAXA-EBIT to monitor the incident photon beam flux.
When the photon beam is incident on the Au foil, a photocurrent proportional to the beam flux is generated. We inserted the photoelectron meter onto the beam axis before and after performing each scan and recorded the photon beam flux.

\begin{figure}[ht]
  \centering
  \includegraphics[width=8.0cm]{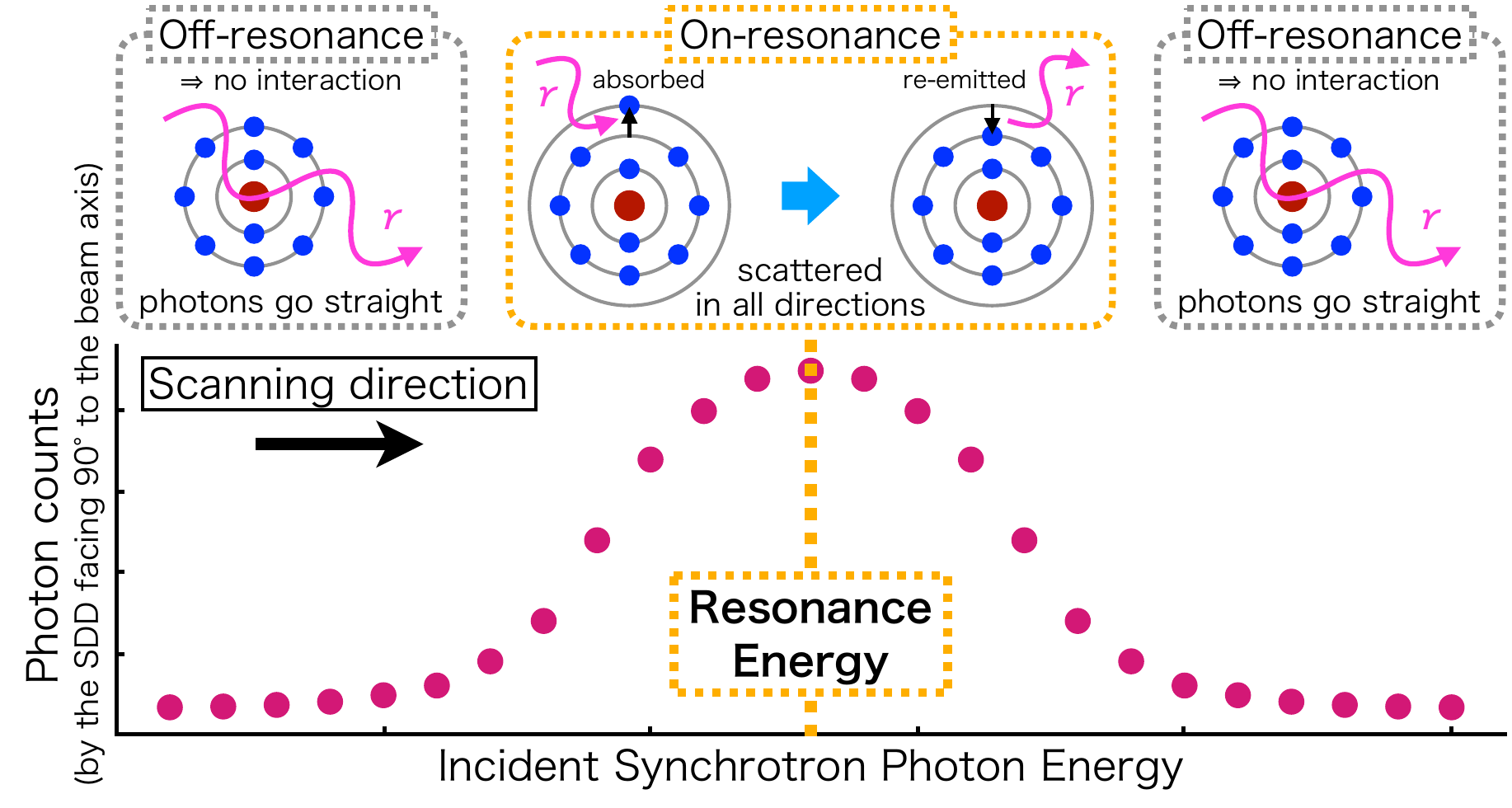}
  \caption{Schematic diagram of the resonant photoexcitation process (top) and the conceptual drawing of the spectrum obtained by scanning incident photon energy (bottom). The scattered photon has the same energy as the incident photon.}
  \label{fig:sync_ener_scan_diagram}
\end{figure}

In this experiment, highly charged iron ions including Ne-like Fe$^{16+}$, the HCI of interest, as well as highly charged oxygen ions including He-like O$^{6+}$, which is utilized in the beamline alignment process described in the next section, were produced by the electron-impact ionization of neutral atoms at the EBIT trap region. 
In our experimental setup, neutral iron was injected into the trap region as a molecular gas of ferrocene (Fe(C$_5$H$_5$)$_2$), while neutral oxygen was inherently present in the EBIT chamber as residual gas including H$_2$O, O$_2$, and CO$_2$. The molecules are dissociated into atoms once they interact with the electron beam.

\subsection{Beamline alignment and optimization}
In the first step of our experiment, we carefully aligned the JAXA-EBIT to overlap the incident photon beam with the ion cloud trapped in the EBIT.
First, we roughly adjusted the position of the EBIT by visually checking the position of the synchrotron photon beam on the YAG screen installed at the downstream end of the EBIT vacuum chamber (far right in Fig. \ref{fig:SP8_beamline_setup_photo}).
Then we further adjusted the position by maximizing the count rate of the K$\alpha$ photons emitted from the resonantly excited He-like O$^{6+}$, to confirm that the maximum overlap between the HCI cloud and the synchrotron photon beam is achieved. 
We used oxygen He$\alpha$ resonance for alignment because the line emission intensity is sufficiently high and the transition energy is well known; thus it is easy to detect using our experimental system.

After confirming the overlap of the ion cloud and photon beam, we searched for the actual resonance energy of each line by roughly scanning the range of $\pm$ 1.5 eV around the resonance energy of the literature value. 
Due to uncertainties in the synchrotron beam energy calibration, the resonance energy is expected to deviate from the literature value. Based on the measured resonance energy, we optimized the scan ranges, step numbers, and slit widths. 
The energy resolution of the synchrotron radiation beam is determined by the exit-slit width of the monochromator; the energy resolution is increased when the slit width is decreased. However, the decrease in the monochromator slit width leads to the deterioration of the signal-to-noise ratio and the possible disappearance of the overlap between the ion cloud and photon beam. The optimal slit width was determined by gradually narrowing the slit width starting from 200 \unit{\micro\meter}, adjusting the alignment, and performing a scan for each slit width setting.

\subsection{Electron beam operation}
The HCIs are bred through electron-impact ionization in the EBIT trap region. A maximum electron-impact ionization cross section is achieved when the electron energy is about three times higher than the ionization threshold (e.g., \cite{Gillaspy01, Lotz68}). 
When scanning the oxygen He$\alpha$ line, the electron beam energy was fixed at 420 eV, which is selected to be about three times higher than the ionization threshold energy of Li-like O$^{5+}$ ions, 138.1197 eV \cite{Lide04}, to maximize the number of trapped He-like O$^{6+}$ ions. 
This electron energy is below the threshold for direct electron-impact excitation of K-shell electrons in He-like O$^{6+}$ (573.94 eV \cite{NISTASD}), thereby suppressing direct excitation events, a source of background.

For the iron 3C and 3G measurement, a certain kind of electron beam operation was required since the detection of these lines is challenging due to a high level of background in a normal EBIT setup.
Instead of fixing the electron beam energy at a certain value, we applied the same type operation as previous studies \cite{Kuhn22, Shah24PRA} to our experimental system in order to increase the signal-to-noise ratio by maximizing the number of trapped Ne-like Fe$^{16+}$ ions while suppressing the background. 
We set the electron beam energy to 1200 eV to efficiently breed Ne-like Fe$^{16+}$ ions, the target of our measurement, which is approximately 2.5 times the ionization threshold energy of M-shell electrons in Na-like Fe$^{15+}$, 489.256 eV \cite{Lide04}, and lower than the ionization threshold energy of Ne-like Fe$^{16+}$, 1262.7 eV \cite{Thompson01}. 
However, this electron beam energy exceeds the excitation thresholds of various Fe L-shell transitions, such as 3C, 3D, 3G, M2, and 3F. This means that the electron beam of this energy induces a strong X-ray background by direct electron-impact excitation of the L-shell electrons \cite{Shah19}. 
In our experimental system, when the electron beam energy was fixed at 1200 eV, the X-ray events resonantly emitted through photon–ion interactions could not be distinguished from the background events due to the low signal-to-noise ratio. 
In order to reduce the background, we cyclically switched the electron beam energy between the breeding and probing energies in a period of 1 s. During the breeding phases, the electron beam energy of 1200 eV was applied for 500 ms, resulting in the production of a sufficient amount of Ne-like Fe$^{16+}$.
During the probing phases, we lowered the electron beam energy to 250 eV for 500 ms. Previous studies \cite{Shah19, Kuhn22} have demonstrated that the background from dielectronic recombination and direct electron-impact excitation is effectively eliminated at such a low electron beam energy of 250 eV. The electron beam at the probing energy of 250 eV cannot produce Ne-like Fe$^{16+}$, resulting in a continuous depletion of Fe$^{16+}$ due to radiative recombination and charge exchange processes \cite{Grilo21}, as well as the loss of HCIs from the trap \cite{Penetrante91}; thus, only the events from the first 200 ms in the probing phases are extracted (Fig. \ref{fig:phase_folded_lc}) for the analysis explained in the next section.

\begin{figure}[ht]
  \centering
  \includegraphics[width=8.0cm]{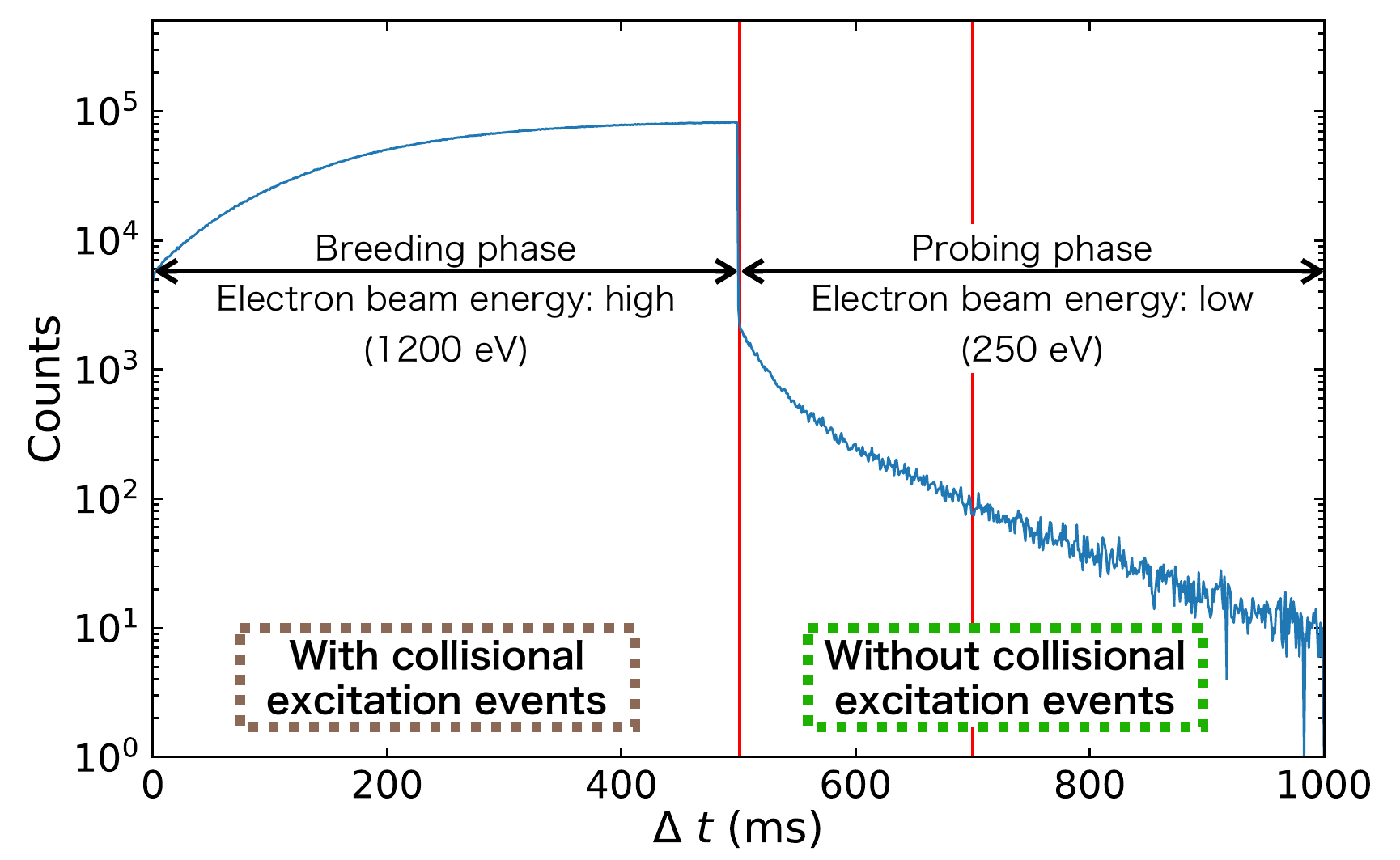}
  \caption{Phase-folded light curve of the detected Fe L$\alpha$ photons within the electron beam energy alternation period of 1 second, summed over the results of 9 scans (total exposure time of 3030 s) performed for 3C measurement.}
  \label{fig:phase_folded_lc}
\end{figure}

The electron beam emission current was 0.61 mA for the oxygen scans, and 2.2 mA during the breeding phases and 1.7 mA during the probing phases for the iron scans.

\section{Analysis and results}

\subsection{Oxygen data}

Figure \ref{fig:O_Hea_result} top panel shows the result of the scan around the resonance energy of the oxygen He$\alpha$ w line. 
A synchrotron photon energy range of 573.63--573.93 eV was scanned in 31 steps, changing the energy 0.01 eV per step.
At each energy step, the photoexcitation signals were integrated for 60 s. 
We find a resonant enhancement of X-ray counts when the incident synchrotron photon energy coincides with the expected resonance energy of O He$\alpha$ w line. 
Note that the pulse heights ($\approx$ 1200 ch.) of the detected resonant events are consistent with that of the collisionally excited O He$\alpha$ photons we measured using the same pulse-height analyzer configuration prior to the beamline experiment.
X-ray events detected within the given region of interest (see Figure \ref{fig:O_Hea_result} top panel) 
are extracted and projected onto the monochromator energy axis. 
The pulse-height range for the event selection is optimized to maximize the signal-to-noise ratio. 
The resulting spectrum is shown in the bottom panel of Figure \ref{fig:O_Hea_result}, where we successfully detect resonance events of the He$\alpha$ w line with a slit width of 50 \unit{\micro\meter}. 
This line was measured for the purpose of aligning the EBIT and beamline, thus a scan with narrower slit width was not performed. 
The spectrum is well reproduced with a Gaussian + linear function model. 
The fit parameters are summarized in Table 1.

\begin{figure}[ht]
  \centering
  \includegraphics[width=8.0cm]{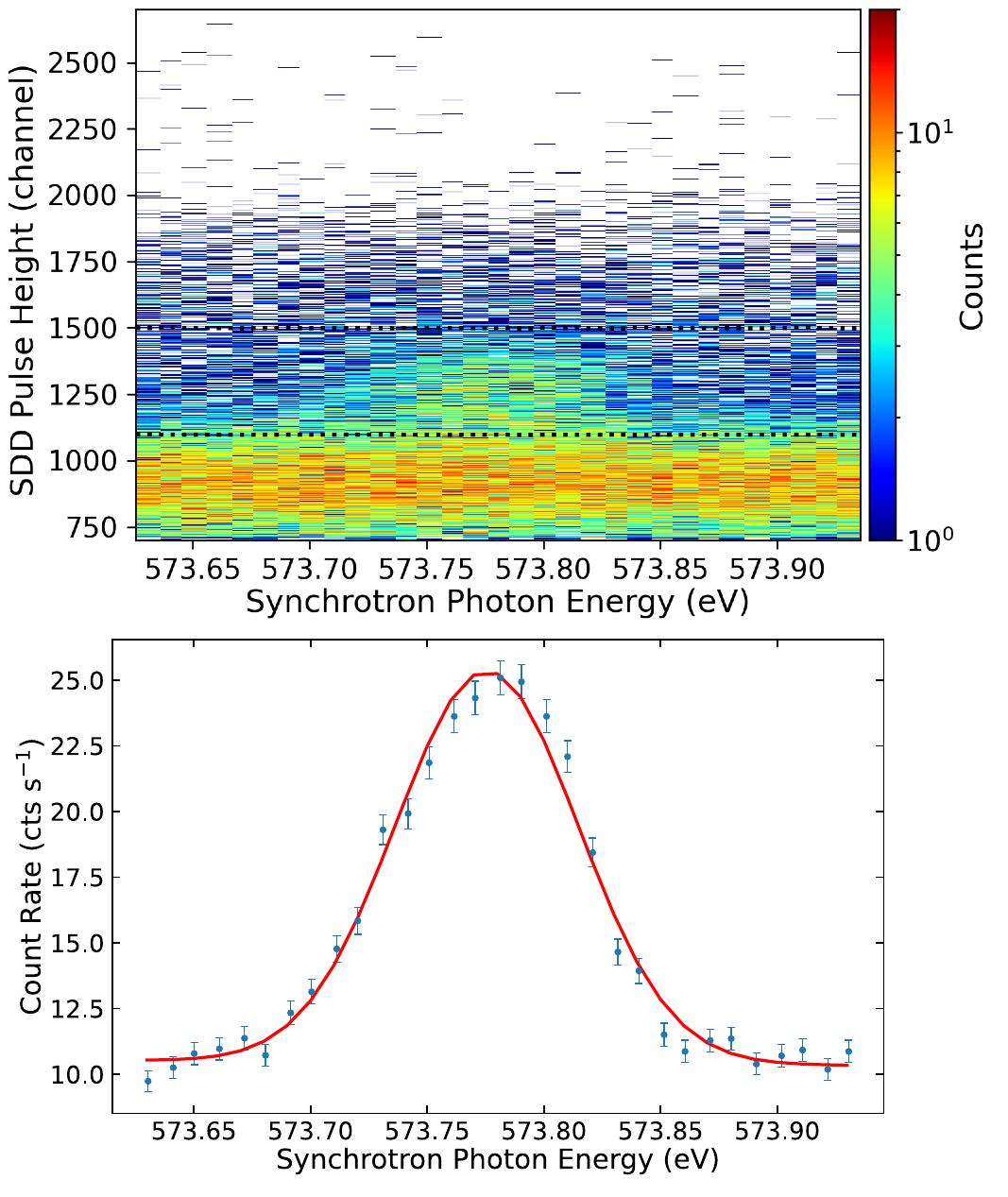}
  \caption{(top) Heat map of photon counts in each pulse-height channel, obtained by scanning the input synchrotron photon energy around the expected centroid energy of O He$\alpha$ w line. 
  The dotted horizontal lines indicate the lower and upper limits of the region of interest for the O He$\alpha$ photons.
  Events registered to this region are selected to plot the projected spectrum shown in the bottom panel.
  (bottom) Count rate of the detected photons for O He$\alpha$ w line in the above-selected region, shown as a function of the input synchrotron photon energy. 3C line is detected and fitted with a Gaussian + linear model. The fit result is shown in a solid red line. }
  \label{fig:O_Hea_result}
\end{figure}

\subsection{Iron data}

\subsubsection{3C line}
Figure \ref{fig:Fe3C3G_results} top-left panel shows the results of the scans around the resonance energy of the Ne-like Fe$^{16+}$ 3C line.  
A synchrotron photon energy range of 825.4--825.9 was scanned in 101 steps, changing the energies 5 meV per step.
At each energy step, the photoexcitation signals were integrated for 30 s.
These scans were repeated 9 times to obtain sufficient statistics in the photon counts. 
In this measurement, we successfully obtained the 3C signals with a slit width of 50 \unit{\micro\meter}. We also attempted measurements with a slit width of 25 \unit{\micro\meter}, but were unable to find a signal.  
We perform the same event selection for the iron 3C data as described above for the oxygen data, with the region of interest determined from the collisional-excitation measurement of Fe L$\alpha$ (see Figure \ref{fig:Fe3C3G_results} top panels). 
As a result, we obtain the projected spectrum shown in the bottom left panel of Figure \ref{fig:Fe3C3G_results}, where we successfully detect resonance events of the 3C line. 
The spectrum is well reproduced with a Gaussian + linear function model, where we could not find the Lorentz wing reported in previous studies (e.g., \cite{Kuhn22}) due to the insufficient signal-to-noise ratio and limited energy resolution of the synchrotron radiation, which is deteriorated by the slit width not being narrow enough.
The signal-to-noise ratio of the obtained spectrum is 0.31, calculated as the ratio of the count rates given by the Gaussian component and the linear-function background component, within the $\pm  1$ sigma range around the central energy of the Gaussian distribution. 
The fit parameters are summarized in Table 1. The area of the Gaussian is determined to be 1.06 $\pm$ 0.06 Counts s$^{-1}$ eV within 6\% precision; a sufficient accuracy is achieved to constrain the oscillator strength ratio within 10\%.

\subsubsection{3G line}
The results of the scans around the Ne-like Fe$^{16+}$ 3G resonance energy and the projected spectrum, extracted from the same region of interest as the 3C data, are shown in the right panels of Figure \ref{fig:Fe3C3G_results}, where we were unable to detect significant resonance events of 3G due to the low signal-to-noise ratio. 
The measured resonance energies of both O He$\alpha$ w and Ne-like Fe$^{16+}$ L$\alpha$ 3C deviate from the literature values (573.94 eV \cite{NISTASD} and 825.83 eV \cite{Loch06} in AtomDB 3.1.3, respectively) by $\sim 0.17$ eV. Since the central energies of these lines are well known, this shift can be interpreted as being due to the systematic errors in our experimental system, for instance, the uncertainty in the energy calibration of the synchrotron radiation. 
The actual resonance energy of 3G in our experimental system is expected to be around 726.90 eV, which is 0.17 eV off the current experimental value of 727.07 eV \cite{Shah24ApJ}.
Therefore, we scanned a synchrotron photon energy range of $\pm$ 1 eV around the expected centroid energy 726.9 eV. 
To be specific, a range of 725.95--727.95 eV 
was scanned in 41 steps, changing the energies 50 meV per step.
The photoexcitation signals were integrated for 30 s at each energy step.
In this measurement, the slit width was set to 100 \unit{\micro\meter} to ensure a moderate energy resolution while gaining a larger photon beam flux, since the 3G line is expected to be much weaker than the 3C due to the oscillator strength of 3G being $\sim 1/20$ of 3C in theoretical calculations \cite{Loch06}.
These scans were repeated 10 times, which was as many scans as we could perform during the beam-time, in attempt to obtain sufficient statistics in the photon counts. 
As with the 3C analysis, we estimated the upper limit of the 3G photon count rate by fitting the data with Gaussian and linear functions. The centrid energy of the Gaussian is fixed at the expected 726.90 eV, and FWHM is fixed at 0.33 eV, which is estimated by scaling the 3C value by the energy and slit width ratio. A 95\% upper limit of the area of the Gaussian is determined to be 0.378 Counts s$^{-1}$ eV.

\begin{figure*}[ht]
  \centering
  \includegraphics[width=16.0cm]{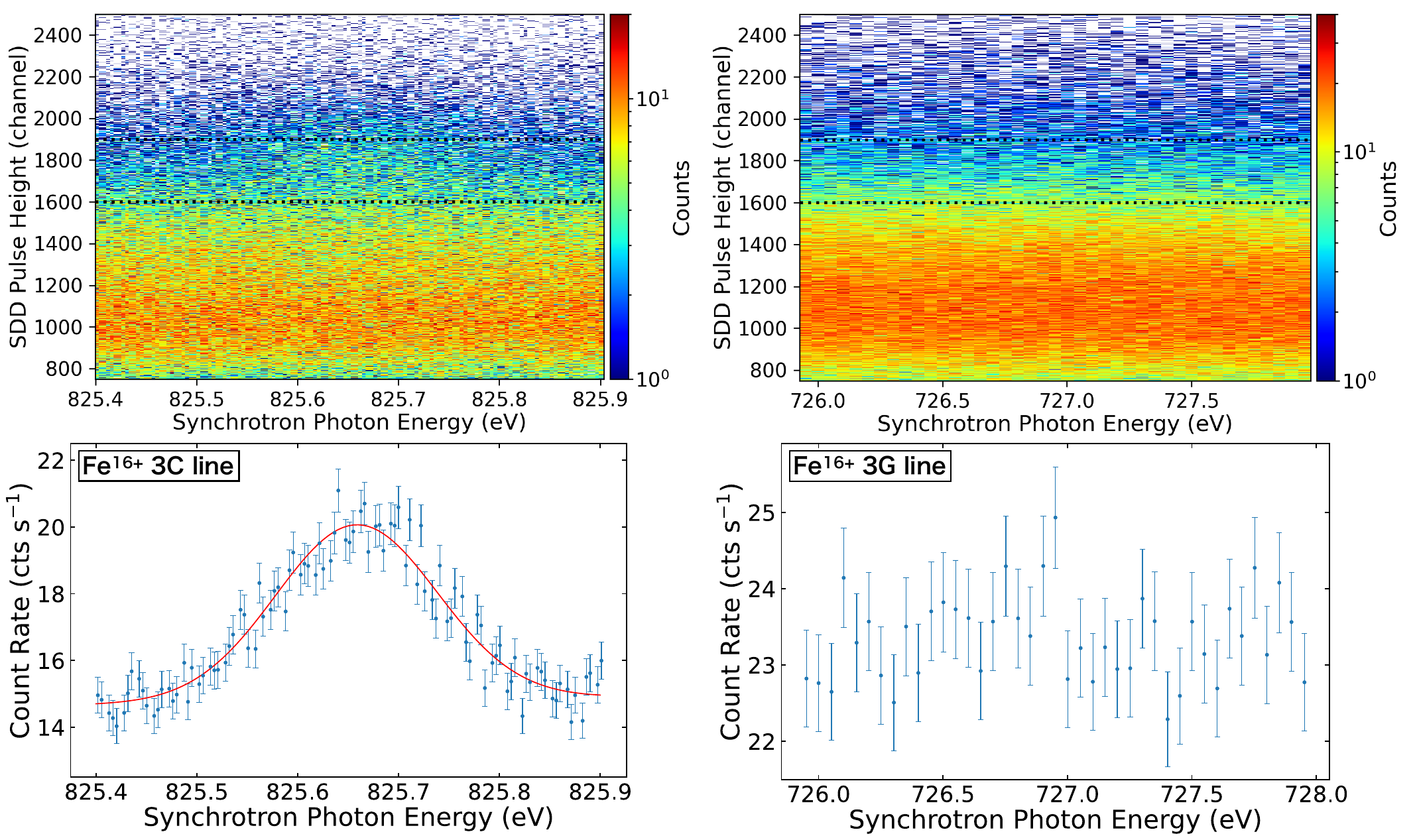}
  \caption{(top) Heat maps of photon counts in each pulse-height channel, obtained by scanning the input synchrotron photon energy around the expected centroid energy of Fe 3C line (left) and 3G line (right). 
  The dotted horizontal lines indicate the lower and upper limits of the region of interest for the Fe L$\alpha$ photons.
  Events registered to this region are selected to plot the projected spectra shown in the bottom panels.
  (bottom) Count rates of the detected photons for Fe 3C (left) and 3G (right) lines in the above-selected region, shown as a function of the input synchrotron photon energy. 3C line is detected and fitted with a Gaussian + linear model. The fit result is shown in a solid red line.}
  \label{fig:Fe3C3G_results}
\end{figure*}

\renewcommand{\arraystretch}{1.5}

\begin{table*}[htb]
  \caption{~ Best-fit parameters for the oxygen He$\alpha$, iron 3C and 3G spectra}
  \label{tab:fit_results}
\centering
  \begin{tabular}{cccc} \hline \hline
     & Centroid energy (eV) & Area (${\rm Counts ~s^{-1} ~eV}$)  & FWHM (eV) \\ \hline
    O He$\alpha$ w & 573.776 $\pm$ 0.001 & 1.46 $\pm$ 0.06 &  0.092 $\pm$ 0.003 \\ 
   Fe 3C & 825.658 $\pm$ 0.003 & 1.06 $\pm$ 0.06 & 0.189 $\pm$ 0.009 \\
   Fe 3G & 726.90 (fixed) & 0.378 (95\% upper limit) &  0.33 (fixed) \\ \hline
  \end{tabular}
\end{table*}

\section{Discussions}

\subsection{centroid energy shift}
The centroid energies, FWHMs and photon count rates of the measured lines are obtained by fitting the spectra with a simple model function.
The wavelengths and oscillator strengths of characteristic X-rays can be constrained using these parameters.
The experimentally obtained line centroid energies directly correspond to the centroid energies of the observed X-ray line spectra, hence precise experimental values can work as benchmarks for the accurate line identification and velocity measurement using Doppler shifts. 
In this experiment, the centroid energy of the Ne-like Fe$^{16+}$ 3C line is constrained to 825.658 $\pm$ 0.003 eV, with a relative precision of $\Delta E/E \sim 4 \times 10^{-6}$, which is significantly better than that of XRISM ($\Delta E/E \sim 8 \times 10^{-4}$). However, we find $\sim$ 0.17 eV shifts from the theoretical values for both the Fe 3C line and the O He$\alpha$ w line that can be interpreted as due to systematic errors in our experimental system, making it difficult to precisely determine the absolute transition energies. 
There can be several factors that cause the shift, one of them being the uncertainty in the energy calibration of the synchrotron radiation. 
Therefore, it is necessary to calibrate the beamline monochrometer with external references. 
One accurate calibration method is to use lines with well-known transition energies, emitted from HCIs inside the EBIT, such as He$\alpha$, He$\beta$, Ly$\alpha$, and Ly$\beta$ lines of various elements. 
State-of-the-art atomic structure codes are capable of calculating these transitions with an uncertainty of $\Delta E/E \sim 10^{-6}$ or less, providing sufficient precision \cite{Leutenegger20}.

\subsection{oscillator strength ratio constraint}
The Gaussian area ratio of 3G to 3C is directly proportional to the oscillator strength ratio. 
By correcting the ratio of the measured photon count rates (obtained as the Gaussian areas) for the ratio of the incident photon beam flux and detector efficiency, a 95\% upper limit of 0.322 is obtained for the oscillator strength ratio $f_\textrm{3G}/f_\textrm{3C}$. 
Considering  the theoretical value of $f_\textrm{3G}/f_\textrm{3C}$, it is possible to detect 3G resonance events and constrain the Gaussian area with the same level of accuracy as 3C, by increasing the signal 20 times or reducing the noise by an order of magnitude. 

The noise that is prominent in our measurement is likely to be the photons of other nearby lines, such as 3C and M2, blended due to the low spectral resolution of the currently used SDD detector ($\Delta E/E \sim 0.02$), and continua, that can be emitted from the recombining processes of the HCIs interacting with the electron beam during the probing phases. 
The noise resulting from the blended lines can be reduced by using a detector with sufficient spectral resolution to separate those lines, for instance, TES microcalorimeters ($\Delta E/E < 5 \times 10^{-4}$; e.g., \cite{Akamatsu09}).
Using time coincidence between photoexcitation events and the pulsed-photon train of synchrotron radiation during data acquisition also helps suppress noise caused by the interaction between the HCIs and the electron beam in general \cite{Bernitt12}.
The synchrotron radiation at the SPring-8 beamlines is generated in the form of pulses of photons with a period of tens to hundreds of nanoseconds and a width of a few picoseconds, depending on the operation mode. Therefore, by reading out the detector events only when photons arrive, noise unrelated to photon irradiation can be reduced.  
Combining these approaches may reduce the noise by an order of magnitude or more.

Another approach is to increase the signals of photoexcitation events 20 times the current level, as mentioned above, and a simple solution is to use a 20 times brighter light source to irradiate the HCIs.
The beam flux of the synchrotron radiation used in this experiment, estimated from the photocurrent measured by the Au foil flux monitor and the quantum efficiency of the Au foil, was $\sim 7.0 \times 10^{10}$ photons s$^{-1}$, therefore incident photon beam flux of more than $1.4 \times 10^{12}$ photons s$^{-1}$ is required for a successful detection of 3G.
Such high beam flux in the soft X-ray energy range will be available at the beamlines of the new-generation synchrotron radiation facilities, such as NanoTerasu (e.g., \cite{Ohtsubo22, Ohtsubo25}) and SPring-8-II (an update project).

\section{Conclusion}

We performed photoexcitation spectroscopy using the JAXA-EBIT at the SPring-8 soft X-ray beamline, to precisely measure atomic data of the astronomically important Ne-like Fe$^{16+}$ L$\alpha$ transitions.
As a result, high-resolution spectra of the He-like O$^{6+}$ $2p$--$1s$ (He$\alpha$ w) and Ne-like Fe$^{16+}$ $\lbrack 2p^5_{1/2} 3d_{3/2} \rbrack_{J=1} \rightarrow \lbrack 2p^6 \rbrack_{J=0}$ (3C) resonance transitions were successfully obtained. 
The Ne-like Fe$^{16+}$ $\lbrack 2p^5_{3/2} 3s_{1/2} \rbrack_{J=1} \rightarrow \lbrack 2p^6 \rbrack_{J=0}$ (3G) resonance transition was also measured, but not detected due to the signal-to-noise ratio being insufficient. 
The oscillator strength ratio of 3G to 3C is experimentally constrained to the 95\% upper limit of 0.322, from these results.
The measured centroid energies of the O He$\alpha$ w and Fe 3C lines are systematically shifted approximately 0.17 eV from the well-known theoretical values.
To detect the weak 3G line and perform accurate measurement of atomic data, such as transition energies and oscillator strengths, we need to improve our experimental system. 
For instance, we needed to improve the signal-to-noise ratio by at least an order of magnitude. 
The noise can be reduced by at least an order of magnitude by upgrading the detector and modifying data acquisition methods.
The signal can be increased by an order of magnitude, for example, by
irradiating the ions with a much brighter light source. The new-generation syncrotron radiation facilities, such as NanoTerasu and SPring-8-II, can be utilized for a follow-up experiment in the near future.
We will apply the experimental values obtained in the present work to observational studies of astrophysical objects to help improve the accuracy of plasma diagnostics using up-to-date high-resolution spectra.
Our future work will also involve comprehensive measurement of different atomic data of astrophysical interest, employing the experimental approach established in this study.


\subsection*{Acknowledgments}
The authors thank Dr. Atsushi Takada and Dr. Takeshi Go Tsuru for their help in the preparation for the present experiment. The presented synchrotron radiation experiment was performed at beamline BL17SU of SPring-8 with the approval of the RIKEN SPring-8 Center (proposal No. 20240064). The authors deeply appreciate all the technical team members of BL17SU, SPring-8. 
This work is supported by JSPS/MEXT Scientific Research Grant Nos. JP24K17106 (Y.A.), JP22H00158 (H.Y.), and JP23H01211 (H.Y.).

\end{normalsize}

\end{document}